\pdfoutput=1
\documentclass[usenatbib]{mnras}

\usepackage{graphicx}	% Including figure files
\usepackage{amsmath}	% Advanced maths commands
\usepackage{amssymb}	% Extra maths symbols
\usepackage[utf8]{inputenc} %Latin Symbols
\usepackage[utf8]{inputenc} %Latin Symbols
\title[The Impact of Redshift on Galaxy Morphometric Classification]{The Impact of Redshift on Galaxy Morphometric Classification: case studies for SDSS, DES, LSST and HST with \textsc{\mdseries Morfometryka}}

\author[Ferreira \& Ferrari]{
Leonardo de Albernaz Ferreira,$^{1}$\thanks{E-mail: leonardo.ferreira.furg@gmail.com}
Fabricio Ferrari,$^{1}$\thanks{E-mail: fabricio.ferrari@furg.br}
\\
$^{1}$Instituto de Matem\'atica Estat\'istica e F\'isica -- IMEF,   Universidade Federal do Rio Grande -- FURG, Rio Grande, RS, Brazil.
}

\date{Accepted 2017 August 30. Received 2017 August 14; in original form 2016 December 29}

\pubyear{2017}

\begin{document}
\label{firstpage}
\pagerange{\pageref{firstpage}--\pageref{lastpage}}
\maketitle
\begin{abstract}
	We have carried  a detailed analysis on the impact of cosmological redshift in the non-parametric approach to automated galaxy morphology classification. 
    We  artificially redshifted  each galaxy from the EFIGI  4458  sample (re-centered at $z\sim 0$) simulating  SDSS, DES, LSST, and HST instruments set-ups
    over the range $0 < z < 1.5$. We then  traced how the morphometry is degraded in each $z$ using \textsc{Morfometryka}. In the process we re-sampled all 
    catalogue to several resolutions and to a diverse signal-to-noise range, allowing us to understand the impact of image sampling and noise on our measurements
    separately. We summarize by exploring the impact of these effects on our capacity to perform automated galaxy supervised morphological classification by 
    investigating the degradation of our classifier's metrics as a function of redshift for each instrument. The overall conclusion is that we can make reliable classification with \textsc{Morfometryka} for $z< 0.2$ with SDSS, for $z<0.5$ with DES, for $z<0.8$ with  LSST and for at least $z < 1.5$ with HST. 
\end{abstract}


\section{Introduction}

The present and near future of extragalactic astronomy will be dominated by multiband photometric imaging data. Following  all developments in extragalactic astronomy and galaxy morphology of the past century (see \cite{Conselice2014} for a review), we are facing an era of large astronomical surveys from both ground-based \citep{SDSSDR8, DES, LSST} and space-based telescopes \citep{COSMOS, JWST, CANDELS, CANDELS2}. Usual classification techniques (i.e. visual classification) are no longer suitable to deal with such a huge volume of data available and even citizen science approaches like Galaxy Zoo \citep{galaxyzoo} are starting to become overrun by the number of objects detected. The ability of distinguishing between galaxy classes is extremely valuable for the study of galaxy formation and evolution \citep{Roberts, bamford2008}, since galactic structure is an ever-changing product of physical processes which happen within such objects and with their environment. The separation of galaxy classes can yield clues and insights not only in the evolution of galaxies but in the evolution of the universe itself.

We still lack a standard method of automated galaxy classification  although many efforts are been made in this direction \citep{Doi1993,odewahn, Huertas2008,MORFOMETRYKA}. In fact, we still do not have a visual robust technique \textcolor{black}{because} the visual classification is  prone to the subjectivity of the investigators \textcolor{black}{ (see fig. 2 of \citet{Fasano2012}) }. The development of such an automated process is highly dependent on our capability to measure structures and features in galaxy images, becoming a difficult task as we increase redshift $z$, since images will be degraded in several aspects. We need to understand not only the impact of image sampling, signal-to-noise ratio (SNR) and photometric depth in morphometry measurements but to know how far it is possible to classify them reliably given an instrumental set-up. In scenarios with low resolution and low SNR (high redshift), the imaging data available would not reflect the galaxy's morphology as observed in high resolution and high SNR. The investigator may be misled when carrying out the same galaxy classification scheme developed using galaxies in the local universe to those galaxies at high redshift whose observations are implicitly very different. For example, \cite{Mortlock2013} show that in the combination of the \cite{Frei} catalogue and a peculiar galaxy sample, up to 50 per cent of the galaxies are misclassified by visual inspectors in redshift simulations, specially disc galaxies since its structure are greatly affected by the loss of resolution. \cite{delgado-serrano} shows that the population of galaxies in the local universe are composed $\sim$70 per cent by spirals while in the distant universe this number drops to $\sim$30 per cent with an increase in peculiars from $\sim$10 per cent to $\sim$50 per cent. Is this difference only due real evolution between these two galaxy populations or is it impacted by misclassification due to lower resolution and SNR? We need to improve our measurements while knowing our limitations. This process is not straightforward since all observational effects are not the same for different instruments.

To make a step forward in the direction of such understanding, we conduct a detailed analysis of the impact of redshift simulations in the automated classification of galaxies, seeing how far can we reproduce the original classification for different instruments in a wide range of redshifts. 
\textcolor{black}{This procedure has been used in some  contexts that are similar to our own. For example, \cite{galaxyzoo2017} apply it in a SDSS galaxy sample to proceed with corrections for redshift-dependent bias in the visual classification of HST data. \cite{BCGFERENGI} uses it to compare local BCGs in SDSS with high-$z$ ones found in CANDELS UDS and \cite{Vika2015} probe how GALFIT \citep{GALFIT2010} measures magnitude while increasing noise and decreasing resolution by simulating galaxies to several redshifts.}

This paper is separated as follows: in \S \ref{ap:degradation} we review how the  cosmology affects the observed sizes and angles; in \S \ref{sec:redproc} we describe all the redshift simulations procedure, data and pre-procssing steps; \S \ref{sec:indexes} defines the measurements used to perform the galaxy classification and how we measure them with the Morfometryka algorithm and how we carry out the supervised classification; In section \S \ref{sec:results} delve into our main findings and how the automated classification is degraded as a function of redshift. We wrap everything up with a brief summary in \S \ref{sec:summary}.

\section{Observational Degradation with Cosmological Redshift}
\label{ap:degradation}

Many modern ground optical telescopes can resolve relevant structures of galaxies in the nearby universe. But this is not true when going deeper in the cosmos, for several effects become non-negligible. As distances increases  in a non-linear manner, several sorts of aberrations are introduced in our observations. For example, we receive less light from sources afar, angular sizes change scale with $(1+z)^{-2}$, the wavelength of a photon increases in its flight time shifting a source's Spectral Energy Distribution (SED) and the equivalent size of instrument's PSF increases, making objects too small to resolve. Our aim here is to quantify how these effects impact the morphological classification of galaxies.

\subsection{Distances and Sizes}

The cosmological model gives us all relationships between distances and sizes and enables us to convert angular separations between two points in an image to a physical distance. All the procedures which follows are dependent on the choice of cosmological model parameters. We adopt the concordance cosmology model, i.e. we use a standard flat $\Lambda-CDM$ model with cosmological parameters given by \citet{PlanckCosmo}, as 
\begin{eqnarray}
    H_0 = 63 \rm \ km s^{-1} Mpc^{-1}, \qquad \Omega_m = 0.316, \qquad \Omega_{\Lambda} = 1 - \Omega_m.
\end{eqnarray}
With that in mind, we follow definitions in \citet{hoggDistances}, where comoving distances are estimated (given a redshift) by 
\begin{eqnarray}
 D(z) = \frac{c}{H_0} \int_{0}^{z} \frac{dz'}{\sqrt{\Omega_{\Lambda} + \Omega_{m} (1+z)^{3}}},
\end{eqnarray}
and with this definition one can measure angular separations in an image as a function of the redshift. The angular diameter distance, $A(z)$, can be written as the ratio between an object characteristic size, $r$,  and its angular correspondence (in radians) in an image
\begin{eqnarray}
	A(z) = \frac{r}{\delta \theta}
\end{eqnarray}
where $\delta \theta$ is the object's extension in radians given the small angle approximation, $\tan(\theta) \sim \theta$. Here $A(z)$ does not have an explicit dependency on $z$, but $z$ is usually necessary to determine $r$. One can also measure the angular diameter distance using a more useful approach using $D(z)$ and $z$ by

\begin{eqnarray}
A(z) = \frac{D(z)}{(1+z)}
\end{eqnarray}
and then the physical size of an object in an image can be estimated by
\begin{eqnarray}
r = \frac{D(z)}{(1+z)} \delta \theta.
\label{eq:angle_to_size}
\end{eqnarray}
\begin{figure}
    \includegraphics[width=\columnwidth]{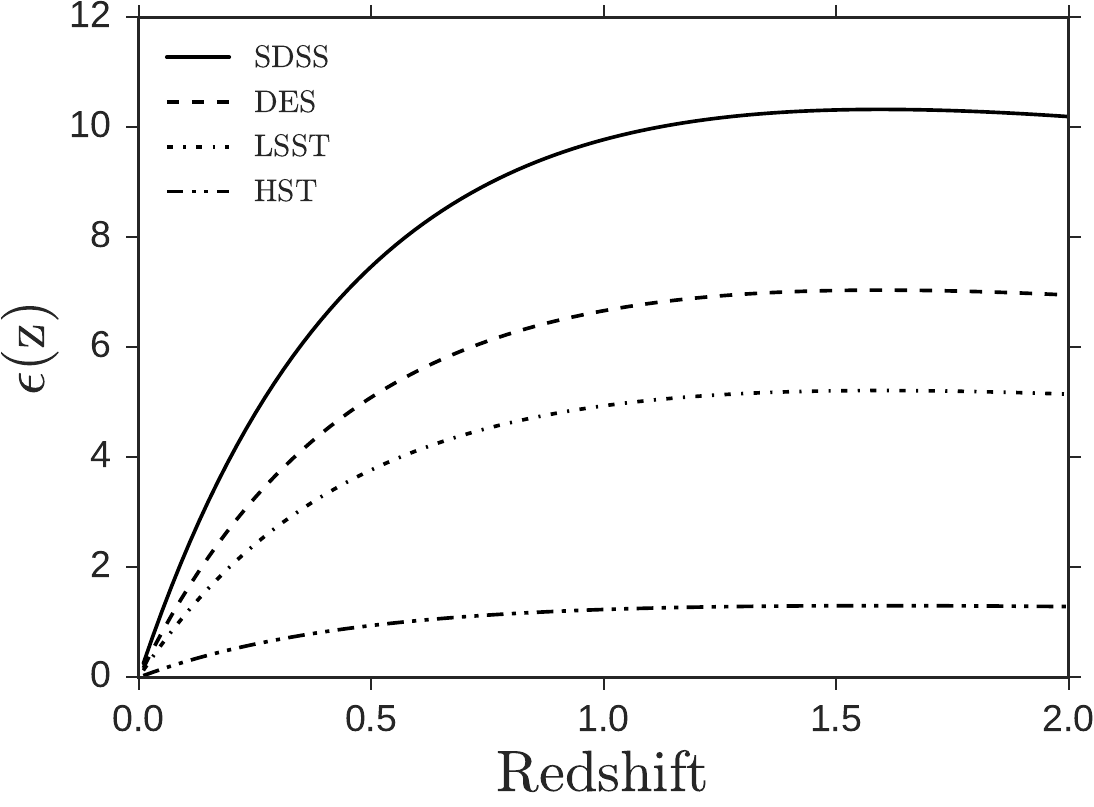}
    \caption{Equivalent size of the smallest resolution element, $\epsilon(z)$ , in kpc for SDSS, DES, LSST and HST. This size is directly related to the instrument's resolution power, lower is better.}
    \label{fig:psfsize}
\end{figure}

\subsection{Point Spread Function}

The Point Spread Function (PSF) of an instrument is \textcolor{black}{the impulse response of the system.} It defines the smallest separation between two point sources in an image that can be resolved individually according to the Rayleigh Criterion \citep{rayleighCriteria}. If the distance within two point sources are smaller than the PSF Full Width at Half-Maximum ($\rm PSF_{\rm FWHM}$), we are only able to detect both sources combined as one. In this way, it is useful to define the $\rm PSF_{\rm FWHM}$ as the smallest resolution element of an astronomical image. The resolving capability of an instrument is degraded as we increase in redshift, since $\rm PSF_{\rm FWHM}$ has a constant size. To estimate this degradation we follow a similar approach to one in \citep{conselice2000A} and we define a resolution element, $\epsilon$, as the ratio between an instrument's $\rm PSF_{\rm FWHM}$ and the angular diameter of $1 \ kpc$ given a distance
\begin{eqnarray}
	\epsilon(z)  = \frac{PSF_{FWHM}}{\theta_{1kpc}(z)}.
\end{eqnarray}
This defines the correspondent size of the PSF in kpc given a redshift and help us to express how much resolution elements is there in a galaxy image. Given the usual PSF's sizes for Sloan Digital Sky Survey (SDSS), Dark Energy Survey (DES), Large Synoptic Survey Telescope (LSST) and Hubble Space Telescope (HST) and its correspondent pixel sizes (Table \ref{tab:simexample}), it is possible to estimate the resolution power of each instrument (Figure \ref{fig:psfsize}) as a function of redshift. For example, a 20 kpc galaxy at $z= 1.5$ would have $\sim 2 \epsilon$ for SDSS, 
\textcolor{black}{which means 2 resolution elements across its diameter and $\pi$ resolution elements in the whole galaxy, if seen face-on.}
The same object would have $\sim 4 \epsilon$ across it for DES, $\sim 5\epsilon$ for LSST and $\sim 20 \epsilon$ for HST. The difference is brutal, HST is one order of magnitude better for resolving galaxies at high redshift. That combined with the fact that angular sizes starts to increase past $z\sim1.6$ \citep{papovich:2003} makes the resolution power of HST reasonably enough to spatially resolve galaxies in \textbf{any} redshift, being only limited by other effects such as cosmological dimming and noise. Overall, LSST is better than DES and DES is better than SDSS, but they poorly resolve galaxies at high redshift.

\subsection{Pixel Resolution Scaling}

Related to this increase in the PSF equivalent size, we have a pixel sampling  downscaling associated by increased distances in higher redshifts. For example, if one measures a galaxy with the same size in different redshifts, $z_1$ and $z_2$, the relation between both image sizes can be found using the fact that the size is constant in Equation (\ref{eq:angle_to_size}),
\begin{eqnarray}
\frac{\delta \theta_1 }{\delta \theta_2}  = \frac{D(z_2)}{D(z_1)}  \frac{(1+z_1)}{(1+z_2)}.
\end{eqnarray}
It is possible to even generalize it for observations with different instruments by writing the angles as
\begin{eqnarray}
\delta \theta = \rm n_{p} p_{s}
\end{eqnarray}
where $n_p$ is the number of pixels substending the angle and $p_s$ the pixel scale of the instrument resulting in
\begin{eqnarray}
	\frac{n_1}{n_2} = \frac{d(z_2)}{d(z_1)}  \frac{(1+z_1)}{(1+z_2)} \frac{p_2}{p_1},
\end{eqnarray}
this is the relation used when adjusting the resolution while simulating data observed with an instrument at given redshift to other instruments and redshifts.

\subsection{Cosmological Dimming}

If two galaxies with the same absolute magnitude, $\mathcal{M}$, are observed in different redshifts, the surface brightness measured for the most distant one would be dimmer than the surface brightness of the closest following 
\begin{eqnarray}
I_0 = \frac{I_e}{(1+z)^4},
\end{eqnarray}
where $I_0$ is the observed surface brightness and $I_e$ the surface brightness as it would be observed locally. This effect is due to time dilatation, increase in wavelength and other geometrical effects imposed by the cosmological model \citep{Tolman1930}. In a close inspection, this effect introduces critical selection biases for high-redshift galaxies \citep{Calvi2014}. As we go deeper, we only observe those galaxies compact and bright enough to compensate this degradation. External dim sections of galaxies that have a bright core tend to fade against the background brightness, potentially resulting in a poor representation of the galaxy true structure. Another point raised by the cosmological dimming is that galaxies at high redshift must have passed through tremendous evolution to become like the galaxies in the local universe, that or the galaxy population in higher redshifts does not represent the progenitors of present-day galaxies \citep{Disney2012}

\subsection{Bandpass Shifting (K-Correction)}

We also have to account for the  K-Correction, which is needed because we do not observe the light from sources in the same rest-frame that it was emitted. The consequence of this effect is that the observed SED of a galaxy is redshifted in comparison to its rest-frame SED, making galaxies in different redshifts to present very different SEDs, resulting in apparent magnitude ($m$) and absolute magnitude (${M}$) in different wavebands. Let $y$ be the band of emission  and $x$ the band we observe; in this configuration, the expression for the magnitudes can be written as \citep{Blanton_kcorrection}

\begin{eqnarray}
	m_x = {M}_y + DM + K_{xy} (z),
\end{eqnarray}
where $m_x$ is the apparent magnitude in the $x$ band, ${M}_y$ is the absolute magnitude in the rest-frame band, DM is the distance modulus and $K_{yx}(z)$ is the correction for the bandpass shift due to $z$.  Writing $K_{xy}(z)$  in terms of the  observed and intrinsic SED, respectively $L(\nu_o)$ and $L(\nu_e)$, we have
\begin{eqnarray}
	K_{xy}(z) = - 2.5 \log_{10} \left [ \frac{L(\nu_e)}{L(\nu_o)} \right ] - 2.5 \log_{10} (1+z).
\end{eqnarray}
In the case of observing high redshift galaxies, this procedure is difficult as it involves matching $L(\nu_e)$ based on observed SEDs of local galaxies of the same Hubble type, which implies in assuming that the SEDs of galaxy types are roughly the same in different redshifts \citep{MoBOOK}. We do not have to face this problem while simulating local galaxies to higher redshifts because we have $L(\nu_e)$ estimates from the multi-band imaging available.

\begin{figure}
    \includegraphics[width=0.95\columnwidth]{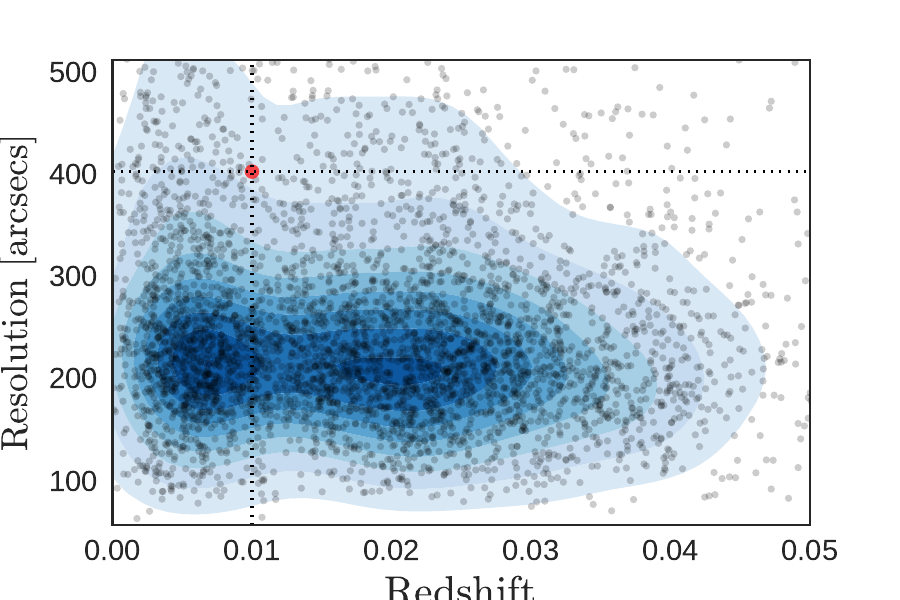}
    \caption{Distribution of redshifts and resolutions for all EFIGI catalog.}
    \label{fig:res_z}
\end{figure}

\section{Redshifting Procedure}
	\label{sec:redproc}

To simulate how a nearby galaxy would appear in higher redshifts we need to account for all effects introduced by the universe dynamics: resolution downscaling, cosmological dimming, bandpass shifting, and K-corrections. We discussed them briefly in Section \ref{ap:degradation}. In general, the investigator must perform predictions for bandpass shifting, downscaling in the image resolution, adjust the flux for the new cosmological distance, apply K-corrections, apply realistic noise and smooth the resulting image by the instrument PSF. A simple recipe for this procedure is given by \cite{conselice2003}, but we choose to adopt a more streamlined approach using the FERENGI algorithm \citep{FERENGI}. We apply the FERENGI algorithm to the EFIGI dataset.

\subsection{The FERENGI algorithm}

Handling all steps aforementioned, FERENGI simulates how a galaxy would be observed in a target redshift given the desired instrument properties. The details of the algorithm are discussed in \cite{FERENGI}. Although FERENGI was developed with the goal of simulating the observation of nearby galaxies in the SDSS into HST surveys like GEMS \citep{gems} and COSMOS \citep{COSMOS}, one can easily adapt this procedure to match its own goal.  Here, we want to simulate the observation of SDSS galaxies in the EFIGI catalog into several different instruments, not only HST. To do that, we made minor changes to the code.
We do not reconstruct the PSF in the way FERENGI does since it does not give desirable results, for our simulations includes low redshift ranges. The direct deconvolution in Fourier space made by FERENGI introduces noise to the resulting PSF when the input and output PSFs are very similar in sizes, which is generally the case for simulating observations from a ground instrument to another. Instead, we choose to use our own PSFs. This not only avoid the deconvolution problem but allow us to use different PSFs that are very different from the characteristic HST PSFs, a essential step since we are interested in analyzing how different instruments observe the same set of galaxies. 
\textcolor{black}{
Besides, we were able to use specific PSF for each setup. 
For SDSS, from where the original images were derived, we have used the appropriate PSF for each 
galaxy and filter. For the other instruments, we have generated PSF according to the  prescription for PSF sizes at each filter.}
Then we created a Python wrapper to facilitate the use of FERENGI to simulate the same galaxy for several target redshifts.

\textcolor{black}{
The galaxy luminosity function is expected to change due to evolution
of galaxies (Willmer et al 2006). \textsc{Ferengi} can account for
such evolution by adding a linear correction $M_{\rm evo} = \alpha z +
M$. Willmet et al 2006 results means at most a decrease of 1 in
absolute magnitude in the range $0<z<1$ (i.e. $\alpha = -1$).  However, in
this work we do not apply such evolutionary corrections, for our goal
is to consider the effects of cosmological distances on the derived
parameters and their effect in the classification.  Indeed, the main
effect degrading the observations is the cosmological dimming (eq
A10), which accounts for a factor 16 in the range $0<z<1$, whilst the
luminosity function evolution would correspond for a factor smaller than
2.5.    
}
    
Another important point is the fact that FERENGI does not perform any source extraction in the input images and adjacent objects could appear as merged with the galaxy when simulated to high redshift. This needs to be handled to avoid that stars and other bright objects impact our results. Pre-processing and segmentation steps are discussed in \ref{sec:prepro} and \ref{sec:segment}. 

\subsection{The EFIGI dataset}\label{sec:data}

In order to understand how observational effects change the way we measure structural parameters in galaxy images we need to have a standard set of galaxy images with detailed morphological information available that is well sampled and has a reasonably good overall SNR. A set of galaxies that fulfill these requirements is the EFIGI catalog \citep{efigiI}, presenting 4458 galaxies in the nearby universe taken from SDSS data. The EFIGI dataset was designed to address the computational aspects of galaxy morphometry. A detailed morphological analysis from several specialists is provided in \cite{efigiII}, giving details about the presence of bars, rings, arms. EFIGI comes with imaging data for the SDSS $ugriz$ bands, resampled for a resolution of  $255\times 255$ pixels$^2$, making each galaxy image to have a different pixel size. Here, instead of using the public imaging data, we choose to use the original stamps from SDSS DR8 \citep{SDSSDR8} frames, provided by the EFIGI team.

\begin{figure}
    \includegraphics[width=\columnwidth]{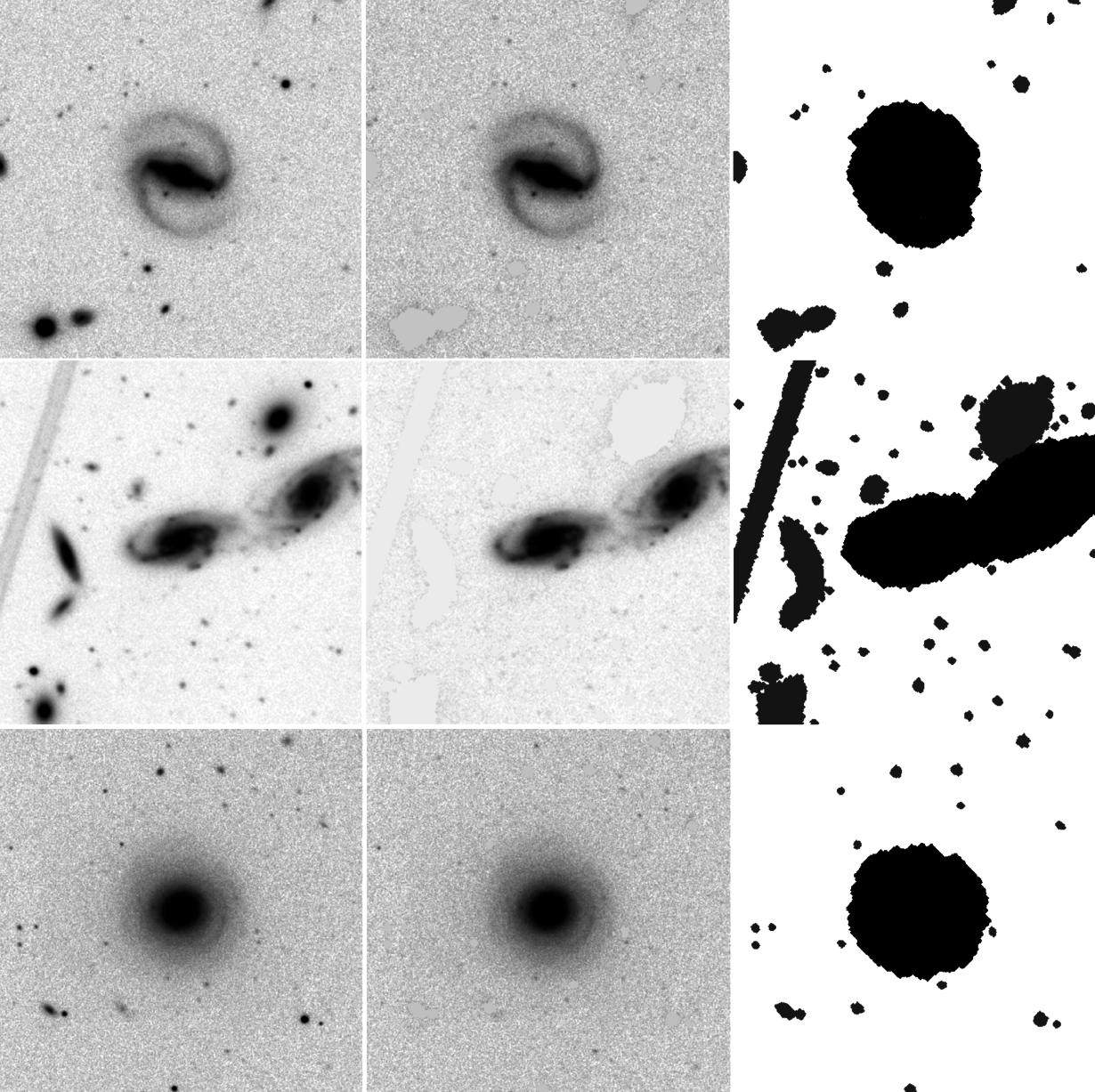}
    \caption{Three examples of the galaxy segmentation procedure. Original images (left), clean images (center) and segmentation map (right). The dark areas are the masks for the detected sources.}
    \label{fig:galaxy_segmentation}
\end{figure}

\subsection{Pre-Processing}
\label{sec:prepro}

The EFIGI galaxies have redshifts in the range  $z = 0.001$ and $z = 0.09$, and they have large variations in image size (Fig. \ref{fig:res_z}). To avoid such variations, we re-scaled all EFIGI galaxies images to the size that they would have if observed at $z = 0.01$; then we re-sampled all images to have the same size. This would again implicate in a pixel scale change for every galaxy, but now we can ensure that this resolution is large enough to sample all galaxies, when they are redshifted, without losing structure information. Since we are at $z = 0.01$ and the SDSS's pixel scale is $0.396 \ \rm arcsec \ \rm pixel^{-1}$, we choose the re-sampling resolution $R_0$ as $400\times 400 \ \rm arcsec^{2}$, making each pixel equivalent to $\sim 80 \ pc$ in physical space\footnote{The cosmological parameters used throughout the paper are given in \ref{ap:degradation}}. That is the equivalent of putting every galaxy of Figure \ref{fig:res_z} on the marked red dot, while there is some galaxies with more pixel resolution than $R_0$, the majority of the distribution is found bellow this limit. Thus, the increase in computation time from using $R_0$ as $800 \times 800 \ \rm arcsec^2$ does not justify its use. With this new sampling, we can think that every galaxy have a new pixel scale or that they have different physical sizes but with the same pixel scale. This is not entirely correct, but as we are only concerned with morphometry, all parameters used in this study are scale independent.

\subsection{Image Segmentation}
\label{sec:segment}

As we go deep in redshift, the angular distance between objects decreases. 
In order to avoid the superposition between the target and field objects we masked the stamps to contain only the target galaxy. 
The procedure follows a background estimation through simple recursive sigma-clipping, then all sources above a flux and spatial threshold are selected and mapped (including the galaxy itself). The image stamps were already prepared in such a way that the main galaxy is located at the center, so to create a non-target  segmentation map we only need to exclude the central detected region that corresponds to the galaxy. Thresholds $f_{\rm lim}$ for an image with pixel values $f_i$ are defined as
 \begin{equation}
 f_{\rm lim} = \widehat{f} + 2\ \widehat{\sigma}_f,
\end{equation}
where  $\widehat{f}$ and $\widehat{\sigma}_f$ are the median and the median absolute deviation (MAD), respectively. Then we select those regions that have at least 1\% of the galaxy image size. The light distribution of the sources may vary with wavelength, and to avoid different galaxy segmentations across bands we use the $r$ image segmentation for all bands, changing only the background segmentation. This ensures that we are using the same region for every filter. The final step in the process replaces all non-target regions with the background median, leaving the overall background distribution the same. A visual example of the segmentation results is found in Fig.~\ref{fig:galaxy_segmentation}.

This segmentation procedure is publicly available in a python implementation under the name \url{galclean}\footnote{\url{https://github.com/astroferreira/galclean}} that makes use of astropy.photutils utilities \citep{astropy}. The simplicity of this segmentation makes it not suitable for crowded fields as there is no de-blending step involved.

\begin{figure}
\centering
    \includegraphics[width=\columnwidth]{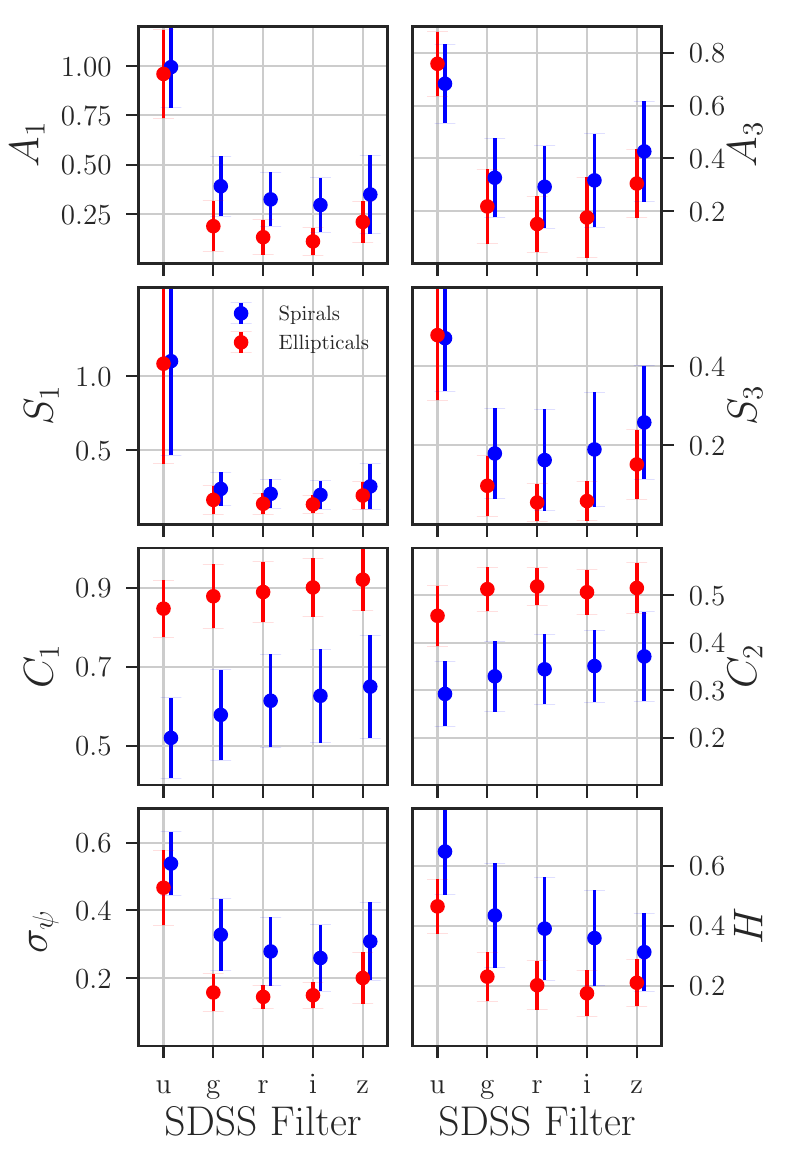}
    \caption{Mean values of $A_1$, $A_3$, $S_1$, $S_3$, $C_1$, $C_2$, $\sigma_\psi$ and $H$ and dispersions ($1 \pm \sigma$) for each band of SDSS ($ugriz$). Colors indicate morphological classes, blue are  spirals and red are ellipticals.}
    \label{fig:bands}
\end{figure}

\begin{figure*}
    \includegraphics[width=0.85\textwidth]{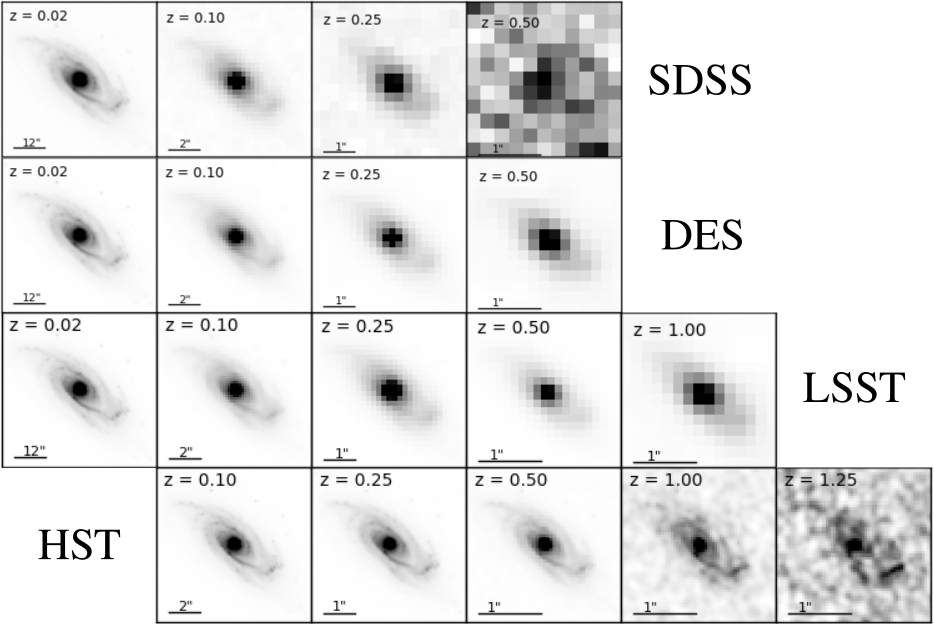}
    \caption{A visual comparison of results of PGC 4992 FERENGI simulations for
     each of the targeted instruments. From top to bottom: SDSS, DES, LSST and HST. The difference in how deep we can resolve structures are evident. }
    \label{fig:simexamples}
\end{figure*}

\subsection{Target Instruments: SDSS, DES, LSST, HST and forthcoming}

The degrees of degradation depend not only on the redshift of the source, but in the instrument configuration as well. To understand in detail how our observations are degraded, we compare how a set of galaxies would appear in instruments with different set-ups. To do that, we choose to conduct this analysis comparing simulated observations from SDSS \citep{SDSSSURVEY}, DES \citep{DES}, LSST \citep{LSST} and Hubble Deep-Field \citep{HDF}. Any instruments not included here that have similar properties to one of these setups should present similar results. The target instruments were chosen because of the following reasons.

The SDSS has defined the standard for wide field galaxy surveys with images and spectra for over 800k galaxies in five bands and was used in the \cite{MORFOMETRYKA} study. Its telescope has a 2.5 m mirror and is located at Apache Point Observatory, in which details are given by \citet{SDSStelescope}. With $\rm PSF_{\rm FWHM} \approx 1.4 \rm \ arcsecs$, pixel size $0.396 \ \rm arcsecs \ {\rm pixel^{-1}}$, it gives us the lower resolution end of all target instruments. As our imaging data is from SDSS and lies outside STRIPE82 \citep{stripe82} region, we choose all redshift simulations to be only single epoch, making SDSS our lower end in depth. 

DES \citep{DES} was chosen because it is a multi-epoch on going wide field survey. With a pixel size of $0.27 \ \rm arcsecs \ pixel^{-1}$, $\rm PSF_{\rm FWHM} \approx 0.7 \rm \ arcsecs$ and a target depth of 10 exposures for each patch of the sky, it presents an improvement over SDSS  both in resolution and depth. The Y band also makes the observation of higher redshift galaxies easier.

LSST  is also included in this work because it will set the standard for wide field surveys in the next decade, mapping 30.000 degree$^2$ of the Southern sky over a 10 years schedule in 6 bands. Its 8.4 m mirror imposes a improvement in photometry depth over SDSS and DES, with a pixel size of $0.20 \rm \ arcsecs  \ pixel^{-1}$ and a $\rm PSF_{\rm FWHM}$  equivalent of that of DES, which is limited by atmosphere. It includes all SDSS bands plus the Y band, with hundreds of planned exposures \citep{LSSTScienceBook} after the 10 year period which will make it the deepest ground wide field survey available.

Finally, the HST is the space instrument chosen to be the best case scenario for this study in terms of resolution. Alongside the lack of atmosphere in the observations, HST with its 2.5m mirror is capable of giving us $0.05 \ \rm arcsecs \ pixel^{-1}$ pixel size, $\rm PSF_{\rm FWHM} \approx 0.15 \ \rm arcsecs$ in general, which is almost an order of magnitude better than the previous instruments. Not only that, but because of its capabilities, HST is responsible for several important surveys like COSMOS \citep{cosmosSurvey} and CANDELS \citep{CANDELS, CANDELS2}. For having several imaging properties and instruments attached to it, we choose to conduct the redshift simulations as it would be observed in the Hubble Deep Field \citep{HDF}, using the same bands, number of exposures and exposure times of the survey, giving also a good depth.

\textcolor{black}{
Morphometry with HST is not limited by angular resolution (\cite{papovich:2003}, their Fig.8) but by SNR. 
Future space telescopes like James Webb Space Telescope (JWST, \cite{JWST:2006})  is expected to go deeper than HST in SNR by a factor
that is roughly given by ratio of their collecting area, if we neglect second order effects such as differences in
efficiency. This factor is given by the ratio between their area, $(4/2.5)^2\simeq 2.6$,  So JWST is expected to go 2.6 deeper in
SNR than HST. Once that cosmological dimming scales with $(1+z)^4$ (Eq.(10)), then a 2.6 factor in SNR means a factor of $\sim 0.21$ in $z$. Given the redshift that HST can reach for morphometric classification, JWST is expected to reliably classify galaxies  $1.2$ times deeper in redshift compared to HST.
}

With this set of target instruments, we can span our simulations in several types of wide field survey equipments in the ground and space, with varying resolutions and depths. Table (\ref{tab:simexample}) summarizes all information from each instrument that are important for the redshift simulations.

\begin{table*}
	\centering
   
    \label{tab:simexample}
     \caption{Information for Instruments and Redshift Simulation. Point source mag limits for SDSS, DES and LSST are presented in 5 $\sigma$. HST is in 10 $\sigma$.}
    \begin{tabular}{lccccc}
        \hline
        Instrument & Pixel Scale &  FWHM  & Filters &  \textcolor{black}{AB mag limit} \\
        \hline
        SDSS & 0.396"/pix & 1.4" &  ugriz  & \textcolor{black}{22.3, 23.3, 23.1, 22.5, 20.8} \\
        DES & 0.27"/pix & 0.7" & grizY & \textcolor{black}{22.9, 22.8, 21.0, 21.4, 20.2} \\
        LSST & 0.20"/pix & 0.7" & ugrizY  & \textcolor{black}{26.3, 27.5, 27.7, 27.0, 26.2, 24.9} \\
        HST & 0.05"/pix & 0.15" & {\tiny F300W, F450W, F606W, F814W}  & \textcolor{black}{26.9, 27.8, 28.2, 27.6} \\
        \hline
       
    \end{tabular}
\end{table*}

\section{\textsc{Morfometryka}}
\label{sec:indexes}
\label{sec:mfmtk}

We have used the \textsc{Morfometryka} \citep[hereafter \textsc{MFMTK}]{MORFOMETRYKA} application to perform the morphometric measurements in the images. We briefly describe \textsc{mfmtk} here; please refer to \citet{MORFOMETRYKA} for full details.

\subsection{Morphometric parameters}
\textsc{Morfometryka}  uses an extended set of parameters composed of the traditional CASGM morphometric system \citep{Abraham1994, abraham1996, conselice2000A, lotz2004}. together with the new parameters  entropy $H$ and spirality $\sigma_\psi$. \textsc{Mfmtk} takes as input the galaxy stamp image and the related PSF, then segmentates it and  measure basic geometric parameters (e.g.  center, axis length, position angle). Next, it quantifies radial light distribution $I ( R )$ from which the Petrosian Radius $R_p$ is defined. For subsequent measurements,  a  Petrosian Region with the same geometric parameters as the galaxy and with a radius of $2R_p$ is used. From it it measures the concentrations $C_1$ and $C_2$, the asymmetries $A_1$ and $A_3$, the smoothness $S_1$ and $S_3$, the Gini coefficient $G$, the second order momentum $M_{20}$, the image information entropy $H$ and the spirality $\sigma_\psi$.

\subsection{Classification Procedure}
\label{ssec:classification}
The first goal of the morphometric procedure carried by \textsc{mfmtk} was to classify galaxies as ellipticals (E) or spirals (S). Again, full detais are given in \citet{MORFOMETRYKA}. Once all the measurements are done, we proceed to evaluate which set of parameters carries most information regarding classification. This is done using the maximum information coefficient (MIC) and comparing the set of all measurements with the known galaxies classes. Originally it was selected the set $\mathbf{x} = \{C_1, A_3, S_3, H, \sigma_\psi \}$.

Once this is done, we perform a Linear Discriminant Analysis (LDA), a supervised classification method, \textcolor{black}{available in the \textsc{Scikit-Learn} python package \citep{scikitlearn}}. In it, given the input vector $\mathbf{x}$ and the E/S classification for the training set, we find a discriminant function which best separates the two classes in the morphometric parameters space defined by $\mathbf{x}$; for the linear discriminant this function describes a plane. This discriminant function is then applied to the target set and each galaxy class is derived from it. This procedure has achieved classification accuracy, performance and recall better than 90 per cent in three different databases.

\begin{figure}
\centering
\includegraphics[width=0.9\linewidth]{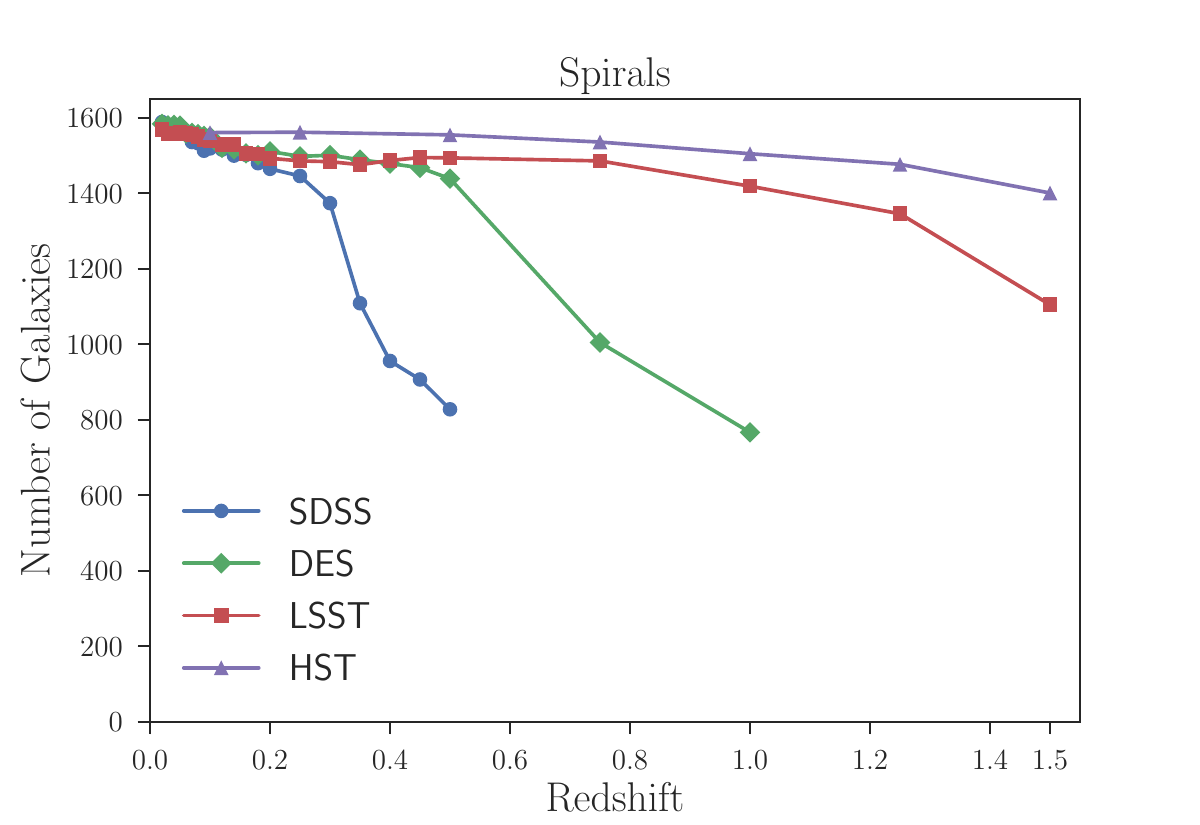} \\
\includegraphics[width=0.9\linewidth]{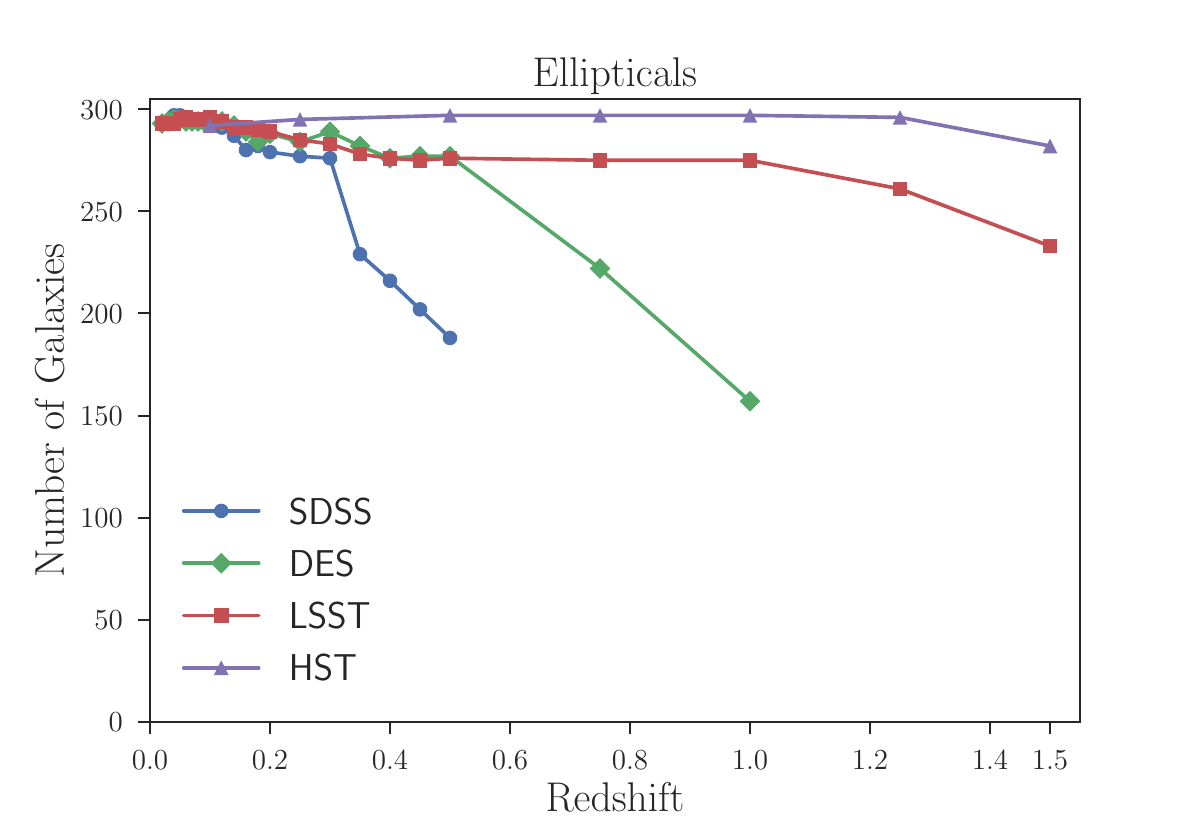}
\caption{\textcolor{black}{The size of the training sample: number of spirals (upper plot) and ellipticals (lower) galaxies with reliable \textsc{Morfometryka} estimates at each redshift which had Galaxy Zoo classification.} }
\label{fig:sample_size}
\end{figure}

\textcolor{black}{
A fraction of galaxies are used to train the classifier, and then the
classifier is applied to the remaining galaxies. To test the classifier we apply a
10-fold cross validation procedure: we use 1/10 of the
sample to train and then classify the remaining 9/10. The procedure is
repeated 10 times such that each of the 10 sets is used as a training
set. The shown values regarding the classifier are mean values for the
10 runs. In our setup we use galaxies classified as elliptical/spiral
in Galaxy Zoo \textcolor{black}{1 data release \citep{galaxyzoo}}.  Figure \ref{fig:sample_size} shows the size of the training sample at each redshift, i.e. how many galaxies, with 
Galaxy Zoo classification, for which it  was possible to apply \textsc{Morfometryka} at each redshift. }

\section{Results and Discussion} \label{sec:results}

Before we explore how the morphometry of nearby galaxies in the EFIGI catalog is altered after redshift simulations and how it changes its supervised classification, we consider some points concerning the parameters and the band-pass shifting of the simulations.  As we are probing the morphometry until $z\approx 1.5$, knowing how it changes with wavelength is specially useful since we will observe these galaxies at  different rest-frame bands. Thus, to check if the morphometry from the original sample at different bands introduce any biases that we would not be aware of, we show in Figure (\ref{fig:bands}) how the morphometry changes with each band from the original data. The $u$-band and the $z$-band have low SNR in general and the impact of noise is important in all parameters except $C_1$ and $C_2$. Because of that, we choose to use the $r$-band to be the target band to be observed in the simulations, since it is one of the bands that shows the best separations between classes in the original data. For each instrument, due to the lack of infrared bands, we will only be able to observe the $r$-band to moderated redshifts. While reaching the redshift limit to observe the $r$-band in the galaxy rest-frame, we then proceed to observe in the $g$-band and then to the $u$-band.  Because of that, we expect $A_1$, $A_3$, $S_1$, $S_3$ and $\sigma_\psi$ to be very noisy and not useful regarding the classification process in higher redshifts by default (while observing the u-band). The concentrations $C_1$ and $C_2$ are shown to be very good in separating classes in every band, so we expect it to be the more robust morphometric feature for the classification procedure. Thus, the information entropy $H$ has big variations between the $g$-band and the $u$-band, but its class separation capability appears to stay approximately the same.

\begin{figure*}
    \includegraphics[width=\textwidth]{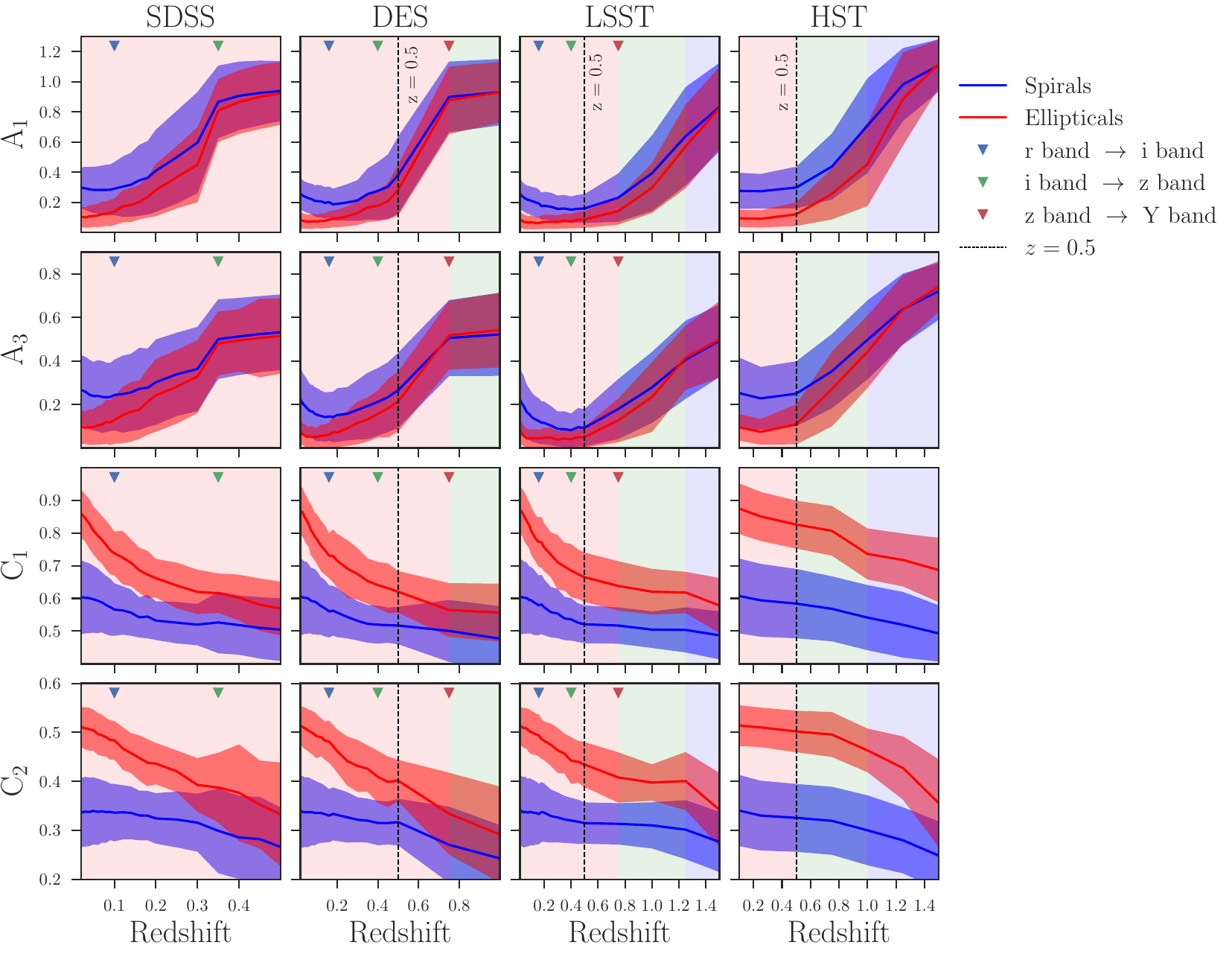}
    \caption{Degradation of $A_1$, $A_3$, $C_1$ and $C_2$. Rows represent morphometric parameters while columns each instrument configuration. Line and filled area colors indicate morphological classes, blue are spirals and red  ellipticals. Solid lines are distribution medians while shaded areas are $1 \pm \sigma$. Background colors indicate which galaxy rest-frame band is being observed in the redshift simulation. Triangles show the redshift steps where the filter in the simulation changes. \textbf{Note the different scales in $z$.}}
    \label{fig:degradation1}
\end{figure*}

\begin{figure*}
    \includegraphics[width=\textwidth]{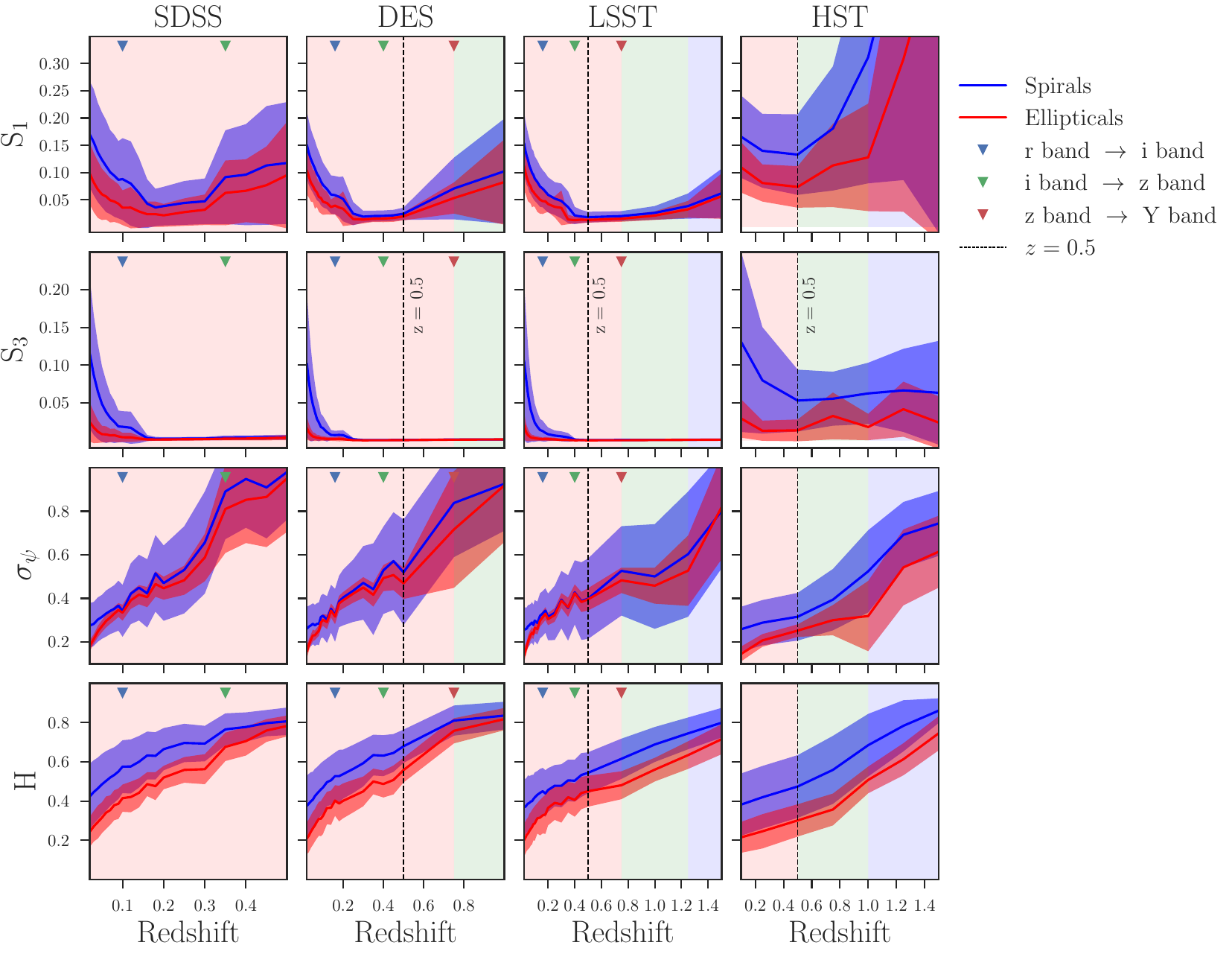}
    \caption{Degradation of $S_1$, $S_3$, $\sigma_\psi$ and $H$. Rows represent morphometry parameters while columns each instrument configuration. Line and filled area colors indicate morphological classes, blue shows measurements for spirals and red measurements for ellipticals. Solid lines are distribution medians while shaded areas are $1 \pm \sigma$. Background colors indicate which galaxy rest-frame band is being observed in the redshift step. Triangles show the redshift steps where the filter in the simulation changes. \textbf{Note the different scales in $z$.}}
    \label{fig:degradation2}
\end{figure*}

We summarize all instrument configurations and redshift simulation ranges for each instrument in Table (\ref{tab:simexample}) and Table (\ref{tab:simrange}), respectively.

\begin{table}
	\centering
    \caption{Simulation Range}
    \label{tab:simrange}
    \begin{tabular}{llr}
        \hline
        Instrument & Range &  \# steps \\
        \hline
        SDSS & $0.02 \le z \le 0.5$ & 20 \\
        DES & $0.02 \le z \le 1$ & 22\\
        LSST & $0.02 \le z \le 1.5$ & 24\\
        HST & $0.10 \le z \le 1.5$ & 7 \\
		\hline
    \end{tabular}
\end{table}

The DES, LSST and HST have multi-epoch imaging. This permits that images taken at different times become combined in one through co-adding processes, improving SNR and depth. As our SDSS sample does not have multi-epoch imaging, we mimic DES, LSST and HST to full depth by stacking FERENGI's simulated images with several Poisson sampled backgrounds to match the number of frames expected to each survey. Thus, we use adequate sky levels for each band of each instrument, improving the SNR of the single epoch simulations. Even not entirely correct, this procedure makes most of the simulated galaxies possible to be detected against the background sky in higher redshifts without the need to use any evolution correction mechanisms or luminosity function corrections, for here we do not want to probe evolution. We leave SDSS simulations as single-epoch observations. This will enable us to show the importance of multi-epoch surveys for the classification even in moderated redshifts. Another important point is that we are not using infrared bands for HST simulations. One should expect HST simulations to present more noise than LSST in higher redshifts, since we are observing the $u$-band in the galaxy rest-frame earlier in redshift than while observing with LSST.

We show examples of the simulation of the galaxy PGC 4992 for each instrument in Figure (\ref{fig:simexamples}). Each line is a redshift step while rows represent the instruments. The power of resolution of small structures is degraded rapidly in the case of ground based instruments, while in the space telescope our simulations are mainly constrained by depth and SNR, not angular resolution. After the simulations and the stacking procedure, all morphometry is then measured for each redshift step for each instrument.

Figures (\ref{fig:degradation1}) and (\ref{fig:degradation2}) shows how $A_1$, $A_3$, $C_1$, $C_2$, $S_1$, $S_3$, $\sigma_\psi$ and $H$ measurements changes with redshift for spirals (blue) and ellipticals (red) separately, where solid lines are distribution medians and solid areas are $1 \pm \sigma$. To distinguish between spirals and ellipticals we use Galaxy Zoo labels \citep{galaxyzoo} as it is the classification standard in automated classification studies. Little triangles in the upper region of the plots represent when the observing band of the simulation is changed while the background colors show the galaxy rest-frame band being observed at given redshift. These transitions are commonly followed by drastic transitions in the distribution of measurements due to the sudden change of sky level associated with the filter being used, resulting in a sudden change of $SNR$. An important point to notice is the redshift ranges of the plots. For SDSS our simulations only reach $z=0.5$, DES reach $z=1$ while $LSST$ and $HST$ reach $z=1.5$. Beyond $z=1.5$ one would need to extrapolate the wavelength limits imposed by the input data, giving non-realistic results.

\subsection{Classification Degradation}

\begin{figure}
    \includegraphics[width=\columnwidth]{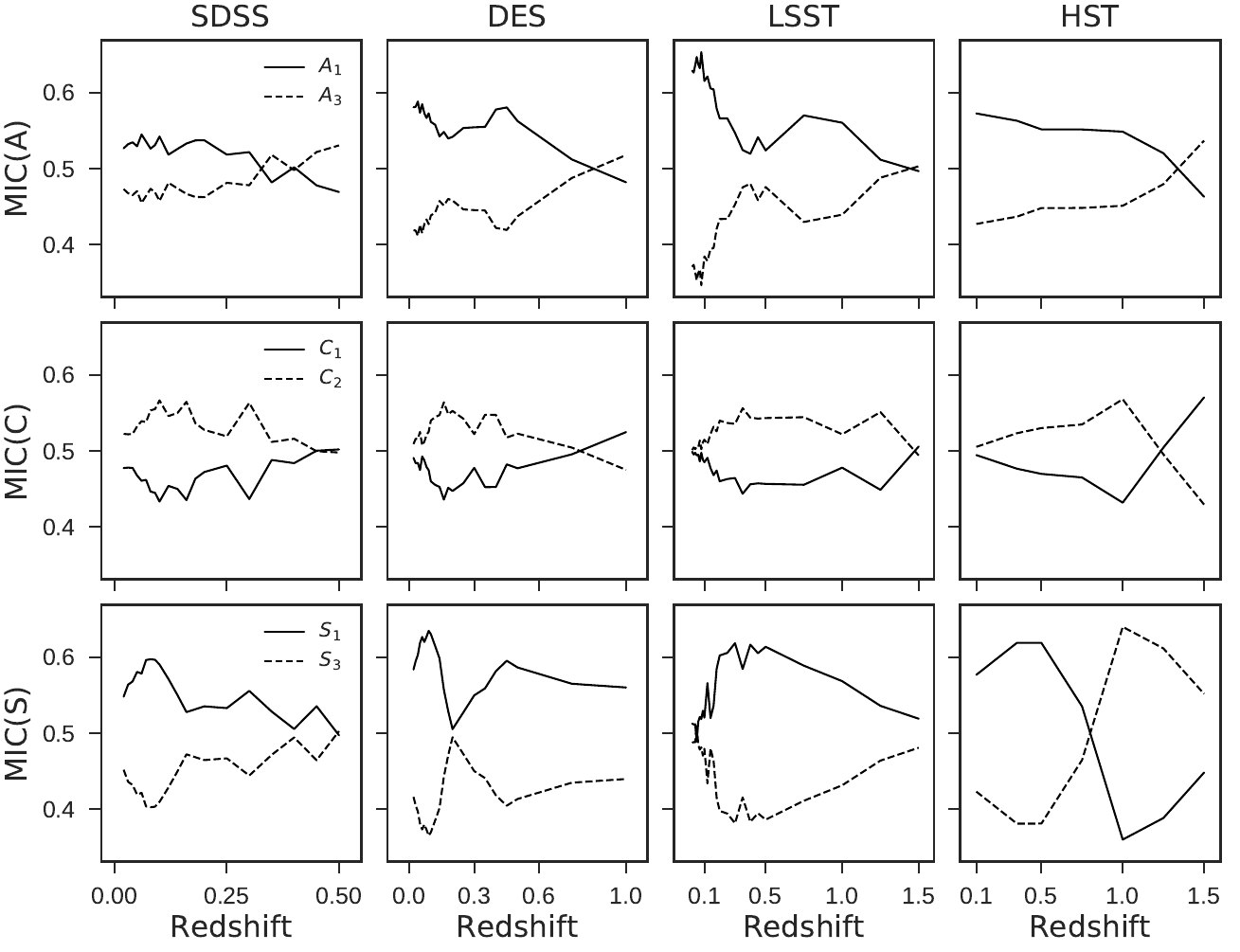}
    \caption{Maximum Information Coefficient (MIC) for each pair of parameters: Asymmetries $A_1$ and $A_3$, Concentrations $C_1$ and $C_2$, and Smoothness $S_1$ and $S_3$. It shows the relative importance of the parameters for the galaxy classification procedure within each pair. Higher is better.}
    \label{fig:MIC}
\end{figure}

We reclassify our sample for each redshift step and for each instrument using the same approach to supervised classification as \textsc{MFMTK}. First, by knowing how our measurements are degraded, we must choose the parameters to the classification that are better in discriminating classes. Since we want a set of independent features, we readily choose $H$. The $\sigma_\psi$ measurement does not show good results as we increase in redshift, so we rule it out of the classification procedure. In the case of the Asymmetries $A_1$ $A_3$, Smoothness $S_1$ $S_3$ and Concentrations $C_1$ $C_2$, we use the MIC \citep{MIC} approach to select the most relevant parameter within pairs. This procedure is also conducted by \textsc{MFMTK}, but in the context of selecting between all parameters. This is done for every redshift step to trace quantitatively which one of the measurements is better to use in the classification procedure. Figure (\ref{fig:MIC}) shows the relative relevance for the classification within pairs, where higher values means more relevant to separate classes. This is important because one could just simply assume from Figure (\ref{fig:degradation1}) and Figure (\ref{fig:degradation2}) which parameters are better without actually quantifying it.

\begin{figure*}
    \includegraphics[width=0.8\textwidth]{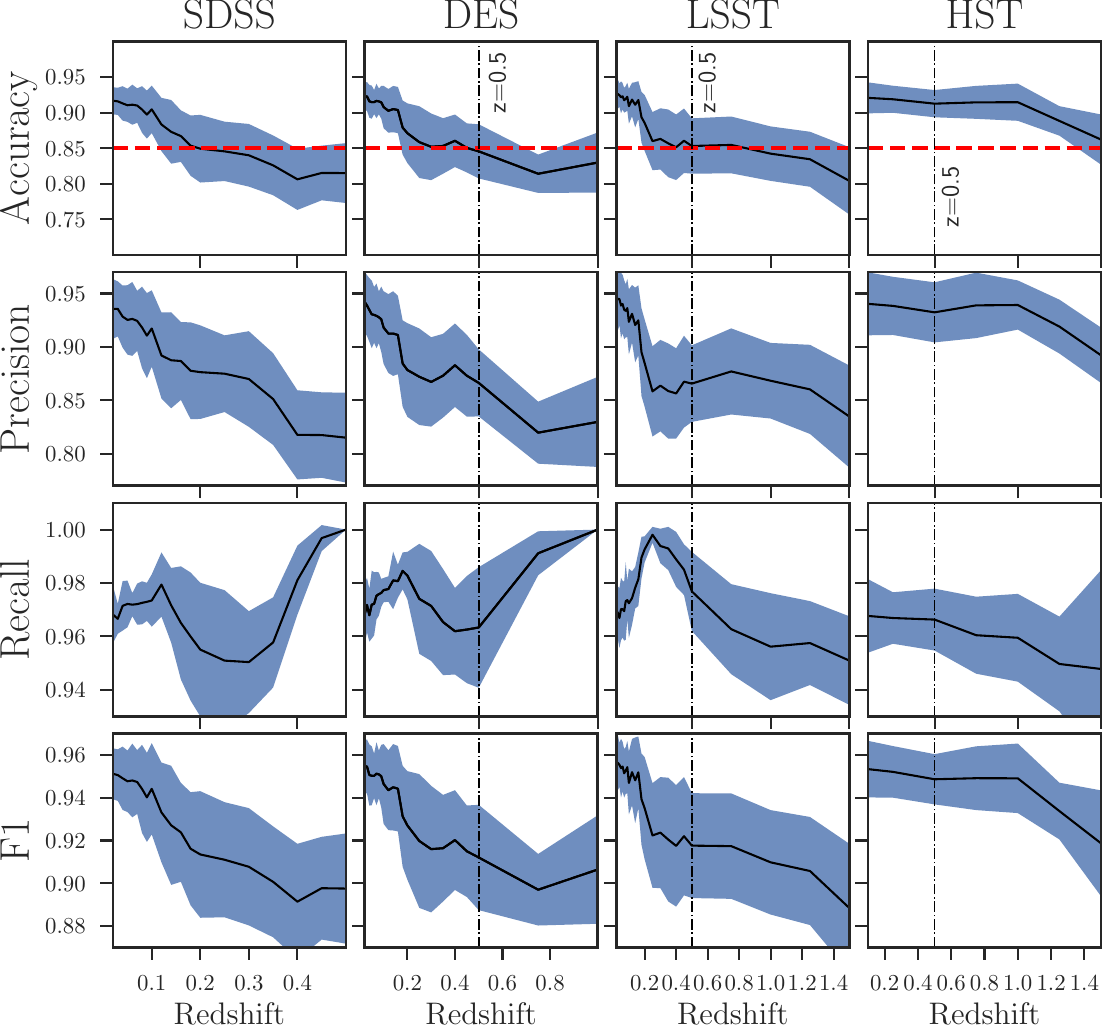}
    \caption{Classification metrics as a function of redshift. Solid lines are medians and shaded area $1 \pm \sigma$. For the Accuracy, we defined the red dashed line $A=0.85$ as a reliability of the classifications. Note that the point dashed line marks $z=0.5$ since the redshift range for each instrument is different.}
    \label{fig:class}
\end{figure*}

For the asymmetry, $A_1$ appears to be better when the data is not noisy, with $A_3$ being more relevant in the tail end of the simulations. In the case of Concentration, $C_2$ is shown to be better throughout all cases, losing relevance to $C_1$ in the noise part of the simulation as well. The smoothness $S_1$ is better than $S_3$ to the separation but shows an odd behavior in the case of HST. We assign it to the fact that for $HST$, $S_1$ diverges when SNR starts to decrease, making the impact of sky background in the measurement much more meaningful for $S_1$ than $S_3$.  We chose the classification features to be
\begin{eqnarray*}
 \mathbf{x} = \{ A_1, C_2, S_1, H  \}.
\end{eqnarray*}
While one might argue that it would be better to use different features for different redshifts, we choose to use a set of independent features that is best in general. 

Now we apply the classification procedure presented in \textsc{MFMTK}  to each redshift step separately. Figure (\ref{fig:class}) shows the classification metrics for SDSS, DES, LSST and HST as a function of redshift. Here we set a reliability threshold for classification as 85\% of the Accuracy, which measures the number of correct classifications within all classifications. In this way, we define the redshift range for the reliability of automated classification in each survey as the range spanning above this threshold. With this scenario in mind, SDSS shows a redshift limit for classification as $z < 0.2$, DES as $z < 0.5$, LSST as $z < 0.8$ while HST does not fall bellow this threshold. This is mostly due to HST's resolution power. This does not mean that we can perform automated morphological classification in HST for any redshift, but that degradation factors are less significant in this case while evolutionary effects would be more relevant in the behavior of morphometric measurements. This same assertive is not true for ground instruments. With evolution effects aside, we can only have confidence in morphological classifications within the imposed limits as degradation effects are much more important in ground-based surveys. To make it even more clear, we explore the impact of resolution and noise in the morphometric parameters in a separate manner.

\subsection{Impact of pixel resolution and noise} \label{sec:res_noise}

Another way of constraining the automated morphological classification process is by defining limits in resolution and noise that will not impact the measurement of the classification features.

\begin{figure*}
    \centering
    \includegraphics[width=0.95\textwidth]{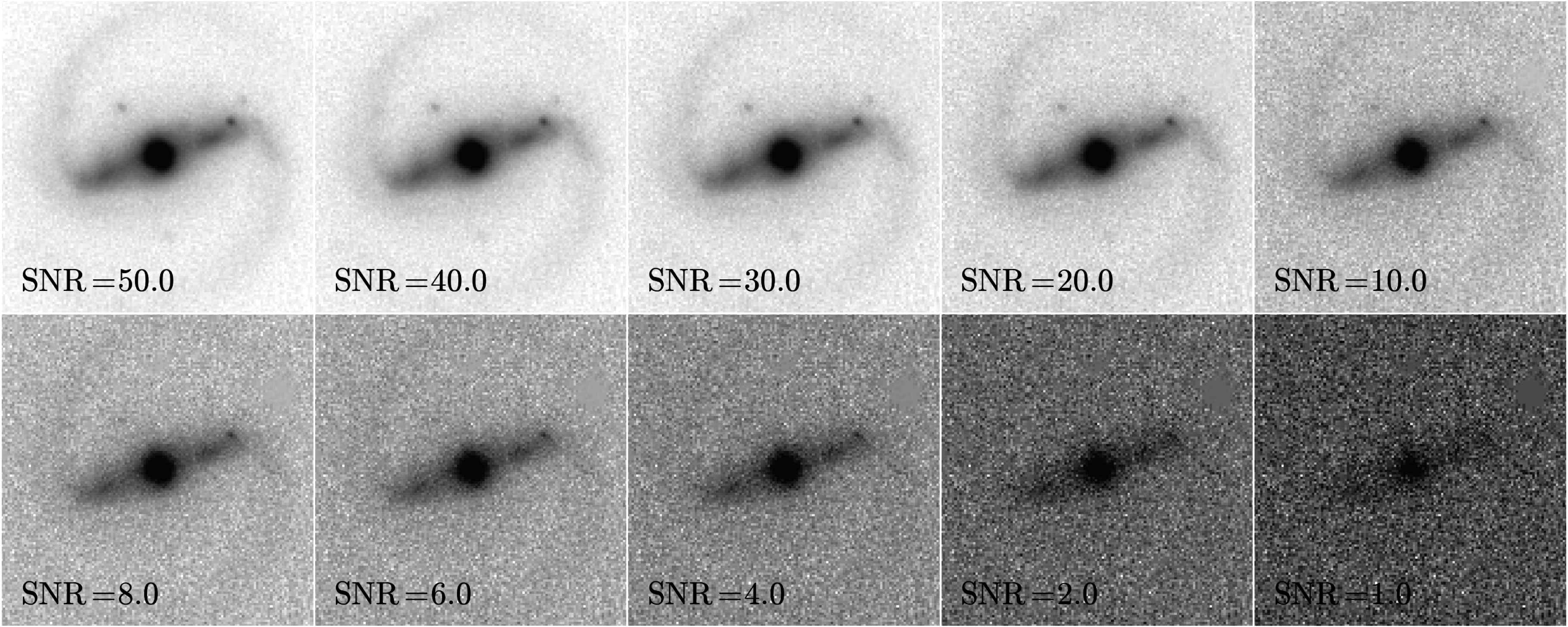}
    \caption{PGC 11670 galaxy for each step of the linear combination to produce the range of desirable SNRs.}
    \label{fig:galaxySNR}
\end{figure*}

First we see how our measurements are affected by noise. With \textsc{Morfometryka} measured properties we define SNR as\footnote{\textcolor{black}{We also did the analysis using $SNR=L_T/(\pi q R_p^2 \sigma_{\rm skybg})$, i.e. the ratio of the mean intensity in the Petrosian region to the sky standard deviation, with similar results.}}
\begin{eqnarray}
	SNR = \frac{I_{2D}}{\sigma_{\rm skybg}}
\end{eqnarray}
where $I_{2D}$ is the effective intensity of a fitted 2D Sérsic profile and $\sigma_{\rm skybg}$ is the sky background standard deviation. Then, we created a sub-sample of EFIGI with all galaxies that had $SNR > 50$ in at least one of the bands, limiting this sub-sample to 284 galaxies. Then, by using a source extracted background from SDSS, we made linear combinations between galaxy and sky background to generate the same galaxy image as it had different SNRs, from 50 to 1. We apply \textsc{Morfometryka} again to the resulting image to certify that it has the desired SNR. A visual example of this process for PGC 11670 is given in Figure (\ref{fig:galaxySNR}). 

\begin{figure}
    \includegraphics[width=\columnwidth]{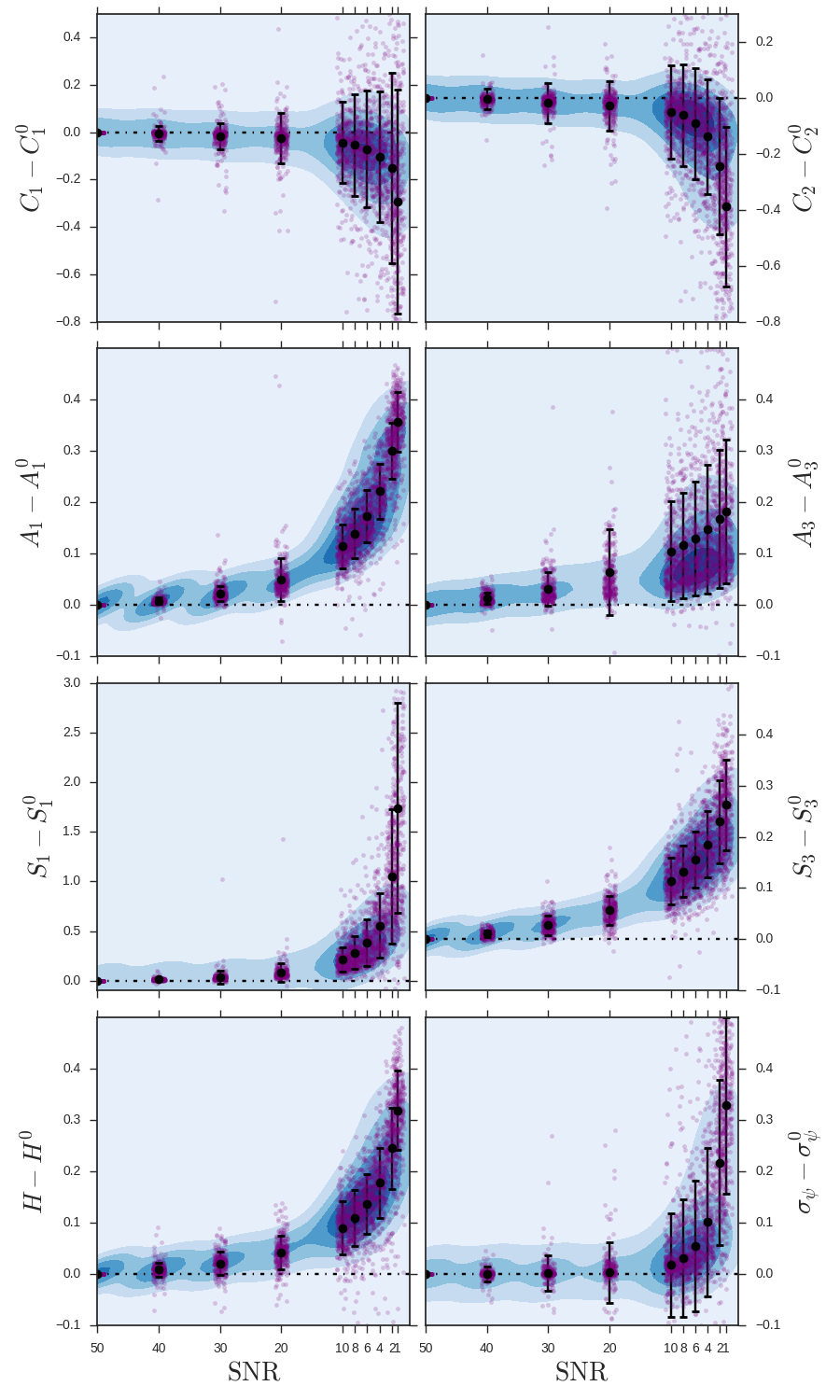}  
    \caption{Deviation from original measurements for $C_1$, $C_2$, $A_1$, $A_3$, $S_1$, $S_3$, $H$ and $\sigma_\psi$. Black circles are medians, error bars are $1 \pm \sigma$ and purple dots jittered data. The background blue density plot shows a KDE estimate for all the distribution of data.}
    \label{fig:graphsSNR}
\end{figure}

The increase of noise makes faint regions of the galaxy fade away rapidly. Figure (\ref{fig:graphsSNR}) shows how $C_1$, $C_2$, $A_1$, $A_3$, $S_1$, $S_3$, $H$ and $\sigma_\psi$ deviates from the original values in the image with $SNR > 50$. Black circles are medians while error bars shows $1 \pm \sigma$. Here, the behavior shown in Figure (\ref{fig:MIC}) is more explicit. For example, $A_3$ and $S_3$ seem to be much more stable in low SNR scenarios, making it more relevant when classificating galaxies with low SNR. In general, all measurements tend to diverge for SNRs lower than $SNR = 10$, and we use this limit as an additional constraint while conducting morphological studies.

\begin{figure}
    \includegraphics[width=\columnwidth]{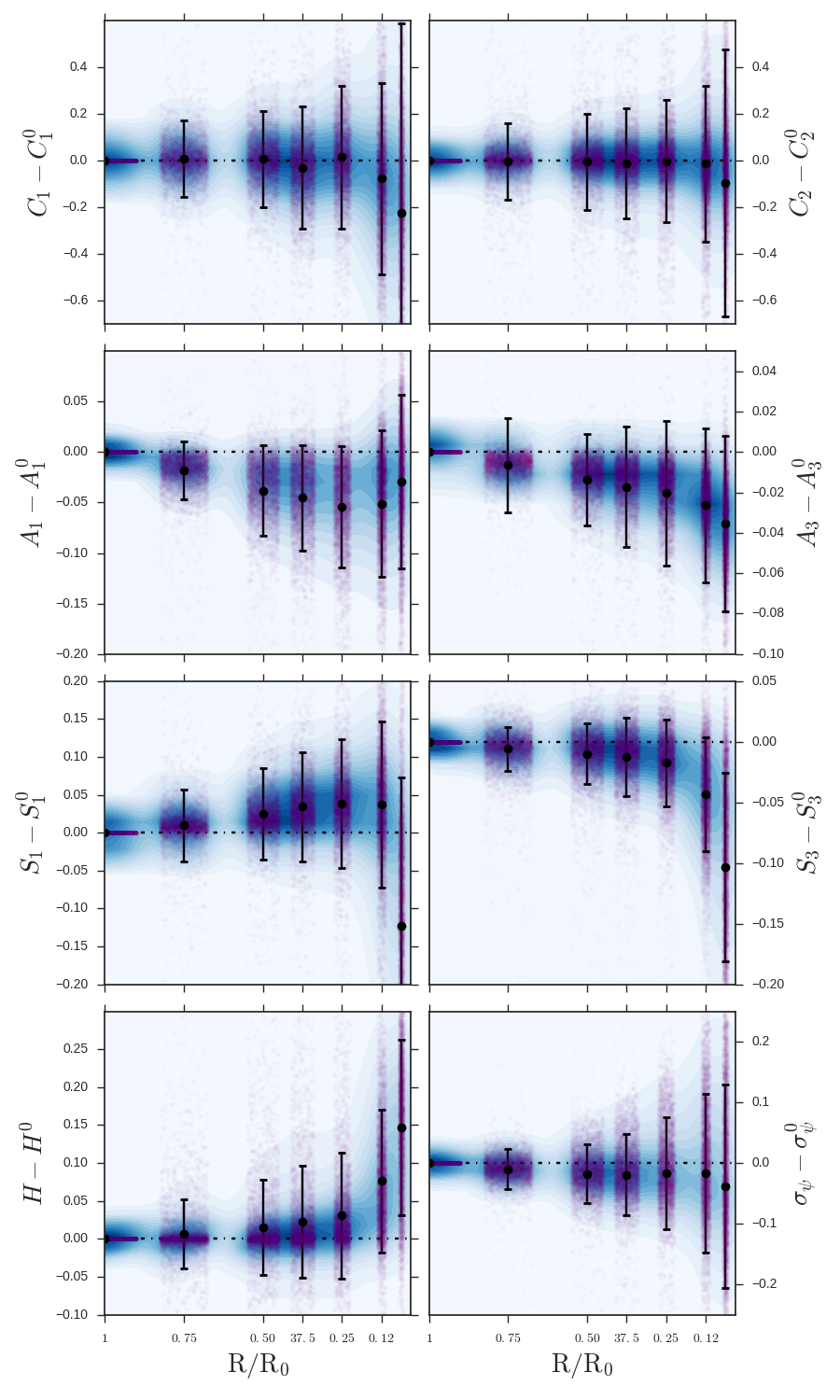}
    \caption{Deviation from original measurements for $C_1$, $C_2$, $A_1$, $A_3$, $S_1$, $S_3$, $H$ and $\sigma_\psi$. Black circles are medians, error bars are $1 \pm \sigma$ and purple dots jittered data. The background blue density plot shows a KDE estimate for all the distribution of data.}
    \label{fig:graphsRES}
\end{figure}

\begin{figure}
    \centering
    \includegraphics[width=\columnwidth]{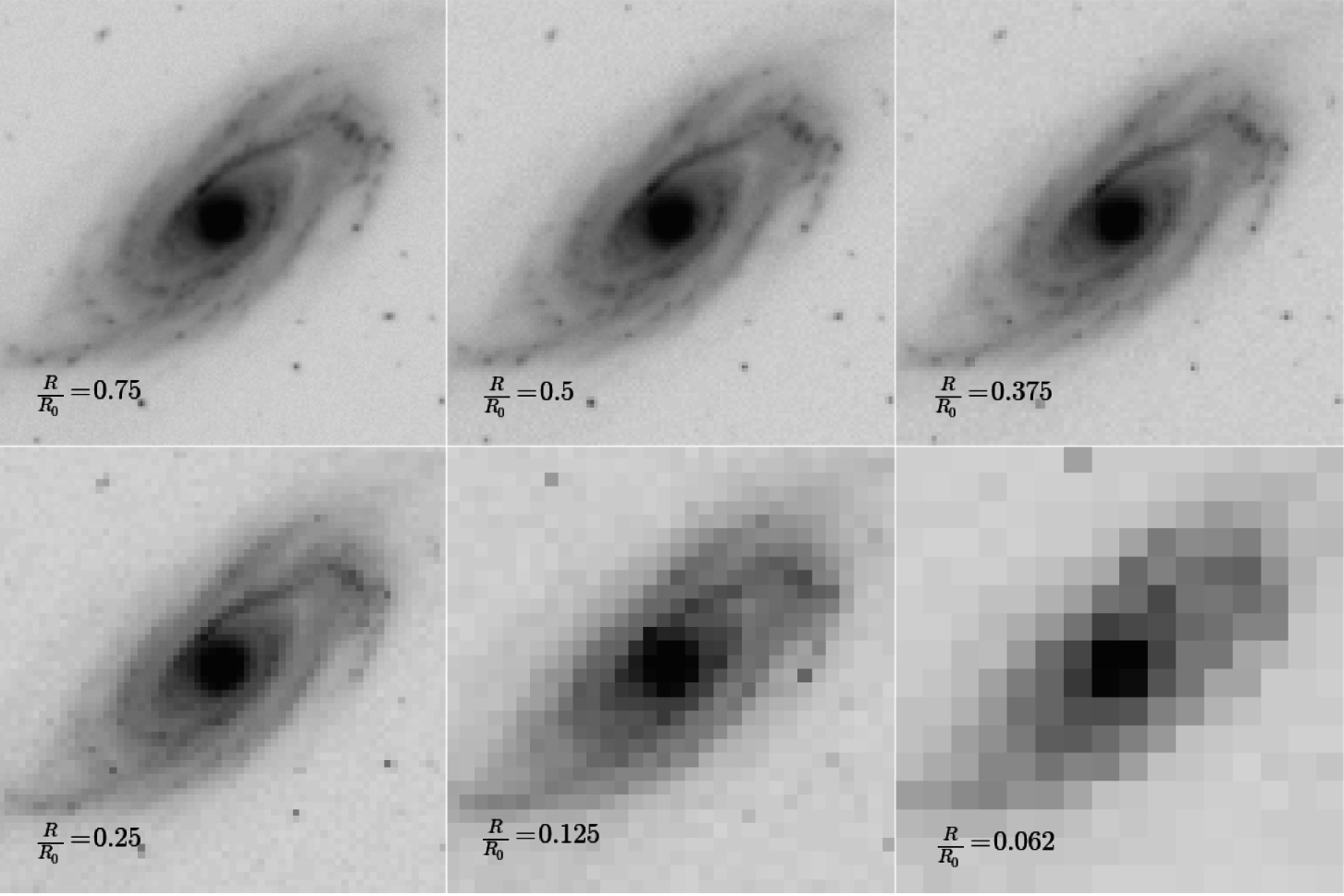}
    \caption{Visual example of the resampling procedure for the PGC 4992 SB galaxy. Visual identification of well resolved spiral arms becomes impossible after critical sampling.}
    \label{fig:resProc}
\end{figure}

A similar approach is taken to verify how the morphometry changes with resolution. All galaxies in EFIGI are resampled to several fractions of the original pixel resolution, namely [0.75, 0.50, 0.375, 0.25, 0.12, 0.065]. Then, we remeasured all morphometry for each of the resolution fractions $R/R_0$, exemplified by Figure (\ref{fig:resProc}).

Given that the original resolution is defined as 400 $\rm arcsec^2$ in $z=0.01$, the re-sampled angular resolution for each step means [4, 2.7, 1.35, 0.67, 0.33]  $arcsec^2$  $kpc^{-1}$ in physical size. The deviations from original measurements are shown in Figure (\ref{fig:graphsRES} in the same way laid out by Figure (\ref{fig:graphsSNR}). It is possible to verify that there is no global trend in the Concentration indexes $C_1$ and $C_2$, only increase in dispersion, which makes them stable with resolution. The Asymmetries $A_1$ and $A_3$ have a tendency to decrease in lower resolutions, which is expected since we expect to resolve less structures that can contribute to an increase in its measurement. This same reasoning could be expected for the Smoothness indexes, but it is only shown in $S_3$. $S_1$ has a trend to go up in lower resolutions. The Information Entropy $H$ increases, as the galaxy image at lower resolutions will have less pixels to represent the galaxy information, thus resembling more a normal distributed image.

\section{Summary}\label{sec:summary} 
We summarize below the important aspects and results concerning the study we carried.

\begin{enumerate}
\item We adapted the FERENGI application to simulate how images are degraded with cosmological redshift for distinct observational set-ups corresponding to SDSS, DES, LSST and HST.
From the EFIGI data base, that was designed for morphological studies,  we simulated how the 4458 galaxies images changed as observed at different redshifts, covering the range $0.01 < z <1.5$.

\item We have used  \textsc{Morfometryka} to investigate how the morphometric parameters change with redshift, applying it to the set of each 4458 galaxies at all the redshifts steps studied.
From this study we could know how each parameter degrades as we move to higher $z$, and how good is its ability to discriminate galaxy classes. The concentrations have shown to be the most stable parameters

\item We also reapplied a supervised classification scheme to the the data set at each different $z$ step to determine how the parameters classification capability was affected. We measure the classifier metrics through accuracy, precision, recall and F1 score.

\item We investigate how the morphometric parameters varied independently with SNR and angular resolution.  For most parameters we can only have reliable results for $SNR>10$.

\item Briefly, we have shown that we can measure morphometric parameters with \textsc{Morfometryka} and achieve reliable classification to $z<0.2$ with SDSS, to $z<0.5$ with DES, to $z<0.8$ with LSST and to at least $z<1.5$ with HST.
\end{enumerate}

\section*{Acknowledgments}
We would like to thank Val\'erie de Lapparent for kindly providing the original EFIGI stamps used in this work. The Programa de P\'os-Gradua\c c\~ao em F\'\i sica of Universidade Federal do Rio Grande for providing the computational infrastructure necessary to conduct all redshift simulations, CAPES and PROPESP-FURG for financial support. We also like to thank the National Science Foundation for travel grants.
%%%%%%%%%%%%%%%%%%%%%%%%%%%%%%%%%%%%%%%%%%%%%%%%%%



\bibliographystyle{mnras}
\bibliography{biblio}

%\newpage
%\listoftodos[Notes]

% Don't change these lines
%\bsp	% typesetting comment

\label{lastpage}
\end{document}